\documentclass[12pt]{article}

\catcode`\@=11

\global\arraycolsep=2pt
\usepackage{amsbsy,amssymb,latexsym,amsfonts,amsmath}
\usepackage{graphicx,color}

\begin{document}
\begin{flushright}
\parbox{4.2cm}
{RUP-19-5}
\end{flushright}

\vspace*{0.7cm}

\begin{center}
{ \Large Conformal equations that are not Virasoro or Weyl invariant}
\vspace*{1.5cm}\\
{Yu Nakayama}
\end{center}
\vspace*{1.0cm}
\begin{center}

Department of Physics, Rikkyo University, Toshima, Tokyo 171-8501, Japan

\vspace{3.8cm}
\end{center}

\begin{abstract}
While the argument by Zamolodchikov and Polchinski suggests global conformal invariance implies Virasoro invariance in two-dimensional unitary conformal field theories with discrete dilatation spectrum, it is not the case in more general situations without these assumptions. We indeed show that almost all the globally conformal invariant differential equations in two dimensions are neither Virasoro invariant nor Weyl invariant. The only exceptions are the higher spin conservation laws, conformal Killing tensor equations and the Laplace equation of a conformal scalar.

\end{abstract}

\thispagestyle{empty} 

\setcounter{page}{0}

\newpage

\section{Introduction}
The Virasoro symmetry is the most powerful as well as ubiquitous symmetries in two dimensional field theories. It is powerful because it allows the classification of critical phenomena from the representation theory of the Virasoro algebra. 
It is ubiquitous because we can show that just the scale invariance together with the unitarity and discreteness of the dilatation spectrum implies Virasoro invariance in two-dimensional quantum field theories \cite{Zamolodchikov:1986gt}\cite{Polchinski:1987dy} (see e.g. \cite{Nakayama:2010zz} for a review). 

Outside of this sanctuary of unitary conformal field theories, however, situations may be more complicated. For example, one may ask if there are any globally conformal invariant field theories that are not Virasoro invariant. At first sight, one might wonder whether it is really possible to write down the condition for the global conformal invariance without the Virasoro invariance even before we talk about their existence. The answer is yes, but it is tricky. In order to realize such an exotic scenario, the trace of the energy-momentum tensor $T_{\mu\nu}$ must be written as
\begin{align}
T^{\mu}_{\mu} = \partial_\mu \partial_\nu L^{\mu\nu} \ , \label{cond1}
\end{align}
where $L^{\mu\nu}$ is a symmetric tensor operator that contains a traceless part, which cannot be reduced to a scalar as $L_{\mu\nu} = \delta_{\mu\nu} L$. Here  unitarity actually ensures the {\it non}-existence of such $L_{\mu\nu}$ due to the unitarity bound, and it is the reason why we rarely see these examples, 
but again outside of the sanctuary of unitary field theories, one may encounter dragons. 

In higher dimensions than two, condition \eqref{cond1} is sufficient for us to improve the energy-momentum tensor so that it becomes traceless and the theory is manifestly conformal invariant. In two-dimensions, however, we cannot improve the energy-momentum tensor by a traceless symmetric tensor because the Ricci tensor satisfies $R_{\mu\nu} = \frac{1}{2} R g_{\mu\nu}$ as an identity, making it impossible to add the improvement term $R_{\mu\nu} L^{\mu\nu}$ to the action. Nevertheless, with condition \eqref{cond1}, the theory is still globally conformal invariant. This impossibility of the improvement of the energy-momentum tensor in two-dimensions leads to the lack of the Virasoro invariance with non-zero $L_{\mu\nu}$.

Are these dragons just imaginary creatures? Not necessarily. A theory of elasticity \cite{Landau}\cite{Riva:2005gd} or a theory of cognition \cite{Bialek:1986it}\cite{Bialek:1987qc}\cite{Nakayama:2010ye} are somewhat realistic and physically appealing examples of such \cite{Nakayama:2016dby}. These theories are effectively described by the fourth-order differential equation given by the second power of the Laplacian:
\begin{align}
\Box^2 \phi ~ = 0
\end{align}
which is globally conformal invariant, but not Virasoro invariant in two-dimensions \cite{Nakayama:2016dby}. One may even speculate that the ``surprising" nature of the success of the Virasoro symmetry in two-dimensions may originate from the lack of the Virasoro symmetry in (the theory of) our cognition. Otherwise, it could have been more intuitive to the human cognition.

Here is a related mathematical question. We often identify the conformal invariance with the Weyl invariance. However, Paneitz pointed out \cite{PA} that the above fourth order differential equation cannot be uplifted to Weyl invariant equations in  two-dimensions. The existence of such obstructions are subjects of mathematical studies in Weyl geometry with great interest over the last couple of decades (see e.g. \cite{math1}\cite{math2}\cite{math3}\cite{GG}\cite{FG}\cite{G} for a partial list). Under which condition are these equations (not) Weyl invariant, and if not, what would be the physical origin of such obstructions?

In this paper, we will discuss the fate of globally conformal invariant differential equations under the Virasoro symmetry and the Weyl symmetry.  We show that almost all the globally conformal equations in two dimensions are neither Virasoro invariant nor Weyl invariant.  The only exceptions are the higher spin conservation laws, conformal Killing tensor equations and the Laplace equation of a conformal scalar. We show a physical origin of these obstructions and give a classification based on the effective action approach.

\section{Globally conformal but not Virasoro invariant differential equations}
In this paper, we are interested in globally conformal differential equations in two-dimensions. For definiteness, what we mean by the global conformal symmetry is the symmetry generated by $P_\mu$ (translation), $J_{\mu\nu}$ (rotation), $D$ (dilatation), $K_\mu$ (special conformal) in two-dimensional Minkowski (or Euclidean) space-time or $L_{\pm 1}$, $L_0$ and $\bar{L}_{\pm 1}$, $\bar{L}_0$ in terms of the conventional Virasoro algebra.  In this section, we are not considering the Weyl symmetry that acts on the metric. We therefore only study (quantum) field theories on flat two-dimensional space-time, and in our language, the (global) conformal symmetry does not change the metric.

In this section, we particularly focus on  the globally conformal invariant differential equations constructed out of Verma module $\mathcal{V}$ of the global conformal algebra:
\begin{align}
\mathcal{V} = \mathrm{span} \{ (L_{-1})^n (\bar{L}_{-1})^{\bar{n}} | h,\bar{h} \rangle : n,\bar{n} \in \mathbb{N} \} \ ,
\end{align}
where $| h,\bar{h} \rangle $ is a (quasi-)primary state annihilated by $L_{1}$ as well as by $\bar{L}_{1}$. $h$ and $\bar{h}$ denote eigenvalues of $L_0$ and $\bar{L}_0$.
As we will see, the generic conformal invariant differential equations are not necessarily constructed out of a single Verma module in non-unitary theories or without the discreteness of the dilatation spectrum, but the constructions based on the Verma module is simple and the most transparent. Since the case with the Verma module becomes building blocks of more general situations, we first discuss the case with the Verma module thoroughly.

For our purpose of classifying the globally conformal invariant differential equations and the Virasoro invariant differential equations, the following observation becomes a key step.
Let us consider the chiral action
\begin{align}
S = \int dz \chi(z) \partial^n \phi(z) \ , 
\end{align}
where $\partial$ is a derivative with respect to $z$, which is identified with $L_{-1}$, and study the variation under the infinitesimal conformal transformation
\begin{align}
\delta \chi(z) &= \frac{p}{2} (\partial \epsilon(z)) \chi(z) + \epsilon(z) (\partial \chi(z)) \cr
\delta \phi(z) &= \frac{q}{2} (\partial \epsilon(z)) \phi(z) + \epsilon(z) (\partial \phi(z)) \ ,
\end{align}
where $p$ and $q$ are conformal weights of $\phi$ and $\chi$ to be determined.
After the integration by part, the variation is given by
\begin{align}
\delta S = \int dz  (q-1+p+n) (\partial \epsilon(z)) \chi(z) \partial^n \phi(z) + n(p+\frac{n-1}{2}) (\partial^2 \epsilon(z)) \chi(z) \partial^{n-1}\phi(z) \cr
+ \frac{n(n-1)}{2}(p+\frac{n-2}{3}) (\partial^3 \epsilon(z)) \chi(z) \partial^{n-2}\phi(z) + \cdots ,
\end{align}
where the omitted terms contain higher derivatives on $\epsilon(z)$.

We see that when $n=0$, if we take $p + q =1$, the variation vanishes for arbitrary $\epsilon(z)$. When $n=1$, if we take $p = q = 0$, the variation again vanishes for arbitrary $\epsilon(z)$. For $n\ge 2$, if we take $p = q = \frac{1-n}{2}$, one may cancel the variation of $\partial \epsilon(z)$ and $\partial^2 \epsilon(z)$, but the full invariance requires $\partial^3 \epsilon(z) = 0$. In other words, for generic $n \ge 2$, it is only invariant under the global conformal symmetry of $\epsilon(z) = l_{-1} + l_{0} z + l_{1} z^2$ rather than the full Virasoro symmetry of arbitrary $\epsilon(z)$ as in  the case of $n=0$ and $n=1$.

Given the above observation, we can easily list and classify the globally conformal differential equations that can be constructed out of the single Verma module. By generalizing the above effective action, we introduce $z$ and $\bar{z}$ and we postulate the effective action
\begin{align}
S = \int dz d\bar{z} \chi(z,\bar{z}) \partial^n \bar{\partial}^{\bar{n}} \phi(z,\bar{z}) \ .
\end{align}
Then we may obtain two equations of motion for $\phi$ and $\chi$. From the above discussions with the assigned conformal weight, when $n$ and $\bar{n}$ are both $0$ or $1$, these equations are Virasoro invariant, but otherwise they are only globally conformal invariant. 

Let us list the Virasoro invariant equations for completeness. For $n=1$ and $\bar{n} = 0$, we have the first order differential equations known as (higher spin chiral) conservation laws
\begin{align}
\partial \phi = 0 
\end{align}
with $h=0$ and arbitrary (half integer) $\bar{h}$. Here we recall the standard notation for the conformal weight $(h,\bar{h})$ with the scaling dimension $\Delta = h + \bar{h}$ and the (Lorentz) spin $J = h-\bar{h}$. Note that in this case, it saturates the chirality unitary bound $\Delta = h+\bar{h} = -J$. 

For $\bar{n}=1$ and $n=0$, we have the similar first order differential equations of the conservation laws
\begin{align}
\bar{\partial} \phi = 0 
\end{align}
with $\bar{h}=0$ and arbitrary (half integer) ${h}$, which saturates the chirality unitary bound $\Delta = h+\bar{h} = J$. 

For $n=1$ and $\bar{n}=1$, we have the Laplace equation
\begin{align}
\partial \bar{\partial} \phi = 0 
\end{align}
with $h=\bar{h}=0$, which is Virasoro invariant. We claim these are the only Virasoro invariant differential equations that can be constructed out of a single Verma module. 

All the other equations 
\begin{align}
\partial^n \bar{\partial}^{\bar{n}} \phi = 0 
\end{align}
 with $h = \frac{1-n}{2}$, $\bar{h} = \frac{1-\bar{n}}{2}$ are globally conformal invariant but not Virasoro invariant.

Alternatively one may look at the results in the following way. Suppose we have a (quasi-)primary operator $\Phi$ with the conformal weight $(h,\bar{h})$. Then  the global conformal invariance allows us to impose 
\begin{align}
\partial^{1-2h} \Phi = 0 \cr
\bar{\partial}^{1-2\bar{h}} \Phi = 0 \cr
\partial^{1-2h}\bar{\partial}^{1-2\bar{h}} \Phi = 0 
\end{align}
if $1-2h$ and $1-2\bar{h}$ are positive integers. One does not have to impose the three equations simultaneously, but if we impose two of them, then it corresponds to a doubly degenerate operator. Actually, either of the first two equations automatically imply the third one, so the truly doubly degenerate operators satisfy the first two equations simultaneously.  
The appearance of the doubly degenerate representation is related to the double poles in the conformal block.

In the rest of the section, we will study the connection between our results and the classification of the globally conformal null vectors in higher dimensions \cite{Kos:2014bka}\cite{Penedones:2015aga} (see also \cite{Osborn}\cite{Poland:2018epd} for reviews). Before doing this, we have a couple of comments about the assumption of the single Verma module. First of all, it is obvious that by combining the above constructed differential operators we obtain more general differential equations (or operator relations) with global conformal symmetry (or Virasoro symmetry in special cases)
\begin{align}
\sum_i c_i \partial^{n_i}\bar{\partial}^{\bar{n}_i} \phi_i = 0 \label{confrel}
\end{align}
as long as $n_i + h_i$ and $\bar{n}_i + \bar{h}_i$ are the same in the sum. 

Note that each term in \eqref{confrel} is a null vector of global conformal algebra, so in unitary conformal field theories with the discrete spectrum, we can conclude that the each term must vanish separately. The most typical situation of such is the spin one conserved current. In unitary conformal field theories with the discrete spectrum, the global current conservation
\begin{align}
\partial^\mu J_{\mu} = \partial \bar{J} + \bar{\partial} J = 0 
\end{align}
reduces to the chiral current conservation $\partial \bar{J} = \bar{\partial} J = 0$, but without the assumption of unitarity and the discreteness of the spectrum, it might not be the case. The most familiar example that fails this chiral separation is the target space $O(N)$ Euclidean rotation current $J_\mu = X^I \partial_\mu X^J - X^J \partial_\mu X^I$ in the model of $N$ free non-compact massless scalars $X^I$. Here the assumption of the discrete spectrum (rather than the unitarity) is violated. There are infinitely many such examples in the non-unitary regime.

Another situation is that the operator relation is given by the non-(quasi-)primary operators. For example, the Virasoro null vector equation
\begin{align}
\partial^2 \phi -\frac{2(2h+1)}{3} [\mathcal{L}_{-2}\phi] = 0 
\end{align}
does not belong to the class we studied. In this case, $[\mathcal{L}_{-2}\phi]$ does not transform as a quasi-primary operator, and our classification does not apply.

In \cite{Penedones:2015aga}, the null vectors of the global conformal algebras in the symmetric traceless tensor representations were studied in general $d$ dimensions. Let us compare their results with ours by setting $d=2$ although the representation may be further reducible in two-dimensions. 

There are three types of null vectors in the symmetric traceless tensor representations of global conformal algebra. The first type (Type I) is given by acting derivatives on spin $l$ symmetric tensors in such a way to maximize the resultant spin:
\begin{align}
P_s (\partial_{\nu_1} \cdots \partial_{\nu_{n_A}} J_{\alpha_1  \cdots \alpha_{l}}) = 0 
\end{align}
Here $\Delta_J = 1-l-n_A$, and $P_s$ is the operation that makes the tensor symmetric and traceless.
 In our two-dimensional language, they correspond to
\begin{align}
\partial^{n_A} O = 0 \cr
\bar{\partial}^{n_A} \bar{O} = 0 
\end{align}
where in the former case $h = \frac{1-n_A}{2}$, $\bar{h} = \frac{1-n_A}{2}-l$ and in the latter case, $h = \frac{1-n_A}{2}-l$, $\bar{h} = \frac{1-n_A}{2}$. Only with $n_A= 1$, they are Virasoro invariant, and the equations correspond to the conformal Killing tensor equations. Otherwise they are merely globally conformal invariant. When $n_A >1$, we are tempted to call them partially conformal Killing tensor equations.


The second type (Type II) is given by partially conserved currents by acting derivatives on spin $l$ symmetric tensors in such a way to minimize the resultant spin: 
\begin{align}
\partial^{\nu_1} \cdots \partial^{\nu_{n_A}} J_{\nu_1 \cdots \nu_{n_{A}} \cdots  \nu_l} = 0
\end{align}
Here $\Delta = l + 1- n_A$ with the  bound $n_A \le l$.  In our two-dimensional language, they correspond to
\begin{align}
\partial^{n_A} J = 0 \cr
\bar{\partial}^{n_A} \bar{J} = 0 
\end{align}
where in the former case $h= \frac{1-n_A}{2}$, $\bar{h} = \frac{1-n_A}{2}+ l$, and in the latter case, $h = \frac{1-n_A}{2}+l$, $\bar{h} = \frac{1-n_A}{2}$.

Finally, let us consider Type III, which is given by the conformal powers of Laplacian acting on  spin $l$ symmetric traceless tensors ($\Box^{n_A/2} + \cdots) O_{\mu\nu\cdots} = 0$, where $\Delta = 1 -\frac{n_A}{2}$ for all $l$. On tensor operators, in addition to the powers of  Laplacian $\Box^{n_A/2} = (\partial_\mu \partial^\mu)^{n_A/2}$, there are many distinct scalar $n_A$-th order differential operators acting on them and we have to arrange them so that the operator is globally conformal invariant. The explicit form can be found e.g. in \cite{Penedones:2015aga}, but it is not important here. We will see some examples below.

When $n_A \ge 2l$, the corresponding two-dimensional equations are
\begin{align}
\partial^{\frac{n_A}{2}-l} \bar{\partial}^{\frac{n_A}{2}+l} O = 0 \cr
\partial^{\frac{n_A}{2}+l} \bar{\partial}^{\frac{n_A}{2}-l} \bar{O} = 0 
\end{align}
with $(h_O,\bar{h}_O) = (\frac{1+l-\frac{n_A}{2}}{2},\frac{1-l-\frac{n_A}{2}}{2})$. When $ n_A = 2$ and $l=0$, it is Virasoro invariant. Otherwise, they are only globally conformal invariant.
Finally, when $n_A \le 2l$, the corresponding two-dimensional equations do not exist (while in higher dimensions they give rise to non-trivial conformal invariant equations).

This may sound surprising at first, so let us present some examples here: when $n_A = 2$ $l=1$, we have the second order differential equations acting on a vector field $J_\mu$ as
\begin{align}
\partial^\mu \partial_\mu J_\nu - 2\partial_\nu \partial^\mu J_\mu = 0 \ .
\end{align}
In two-dimensions, it leads to
\begin{align}
\bar{\partial}^2 J &= 0 \cr
{\partial}^2 \bar{J} &=0 \ .
\end{align}
Note that the $J$ value changes its sign by the differentiation but it is a scalar operator in the higher dimensional sense (because they are in the same representations of $SO(d)$). 
On the other hand, when $n_A=2$, $l=2$, the similar would-be conformal second order differential equations on the traceless tensor:
\begin{align}
P_s \left(\partial^\mu \partial_\mu H_{\rho \sigma} - 2 \partial_{\rho} \partial^\mu H_{\sigma\mu } \right)= 0 
\end{align}
holds as an identity in two-dimensions due to the $P_s$ operation that makes the tensor symmetric and traceless.
Thus, there is no corresponding non-trivial conformal invariant differential equation in two-dimensions.

In this way, most of the globally conformal differential equations we constructed can be originated from the conformal null vectors studied in \cite{Kos:2014bka}\cite{Penedones:2015aga}. However,  let us point out that 
\begin{align}
\partial^{n_A} J = 0 \cr
\bar{\partial}^{n_A} \bar{J} = 0 \label{missing}
\end{align}
with $n_A >l$ did not appear in their approach because of the upper bound on the  number of derivatives in type II. Thus, the null vectors constructed out of the symmetric traceless tensor in  \cite{Kos:2014bka}\cite{Penedones:2015aga} cannot explain all the globally conformal invariant equations in two-dimensions. To avoid the confusion, however, let us simply point out that we may always construct such equations by using two-dimensional epsilon tensor. The approach taken in  \cite{Kos:2014bka}\cite{Penedones:2015aga} are applicable to general dimensions, so they did not use the dimension specific epsilon tensor.

This brings us back to the study of the poles in the conformal blocks that are related to the existence of null vectors. Consider the four-point (global) conformal blocks of scalar operators in two-dimensions. By using the (dressed) hypergeometric function
\begin{align}
k_{\beta}(x) =  x^{\beta/2} \ _2F_1 \left(\frac{\beta-\Delta_{12}}{2},\frac{\beta+ \Delta_{34}}{2};\beta;x \right)
\end{align}
they are expressed \cite{Dolan:2003hv} as
\begin{align}
g_{h,\bar{h}}^{\Delta_{12},\Delta_{34}}(z,\bar{z}) =  k_{2h}(z)k_{2\bar{h}}(\bar{z}) +  k_{2h}(\bar{z})k_{2\bar{h}}({z}) \ .
\end{align}
The most general case of external spins can be found in \cite{Osborn:2012vt}.
As we can see, the conformal blocks have poles when either $1-2h$ or $1-2\bar{h}$ is a positive integer, and double poles when both are positive integers. This is in perfect agreement with our results.

However, there remains a small puzzle: what happened to the missing piece \eqref{missing} in the  approach taken in  \cite{Kos:2014bka}\cite{Penedones:2015aga}? Actually, in spite of this missing piece, we can verify that the structure of the poles presented in  \cite{Kos:2014bka}\cite{Penedones:2015aga} is correct in two-dimensions. This is because the location of the poles predicted by the missing piece is precisely the one that we naively expected but did not actually exist in type III, so eventually the missing piece in type II is taken care of by the naive over-counting in type III.

\section{Weyl (non)-invariance}
We now argue that the globally conformal invariant differential equations without Virasoro invariance cannot be uplifted to Weyl invariant equations in curved background. By Weyl invariance, we mean the change of the metric $g_{\mu\nu} \to \Omega^2(x) g_{\mu\nu}$ and dynamical fields $\phi \to \Omega(x)^{-\tilde{\Delta}} \phi$ with the so-called Weyl weight $\tilde{\Delta}$.\footnote{For lower index spin $l$ tensors, the Weyl weight and conformal weight are related by $\Delta = \tilde{\Delta} + l$.}
The above statement is essentially a contraposition of Zumino's theorem \cite{Zumino:1970}. Zumino's theorem claims that the theories that are invariant under diffeomorphism and Weyl transformation are conformally invariant in flat Minkowski (or Euclidean) space-time. In particular, it possesses the full Virasoro invariance.

The proof of Zumino's theorem goes as follows. Let us first notice that the conformal generators are solutions of the conformal Killing equations,
\begin{align}
\partial_\mu \epsilon_\nu + \partial_\nu \epsilon_\mu = \frac{2}{d}\partial^\rho \epsilon_\rho \delta_{\mu\nu} \ ,  \label{kill}
\end{align}
 meaning that they are given by the diffeomorphism whose action can be compensated by the simultaneous Weyl transformation. Then if the theory is invariant under the diffeomorphism and the Weyl transformation, the combined action acts only on the dynamical fields rather than the metric. This implies that this combined action, which is nothing but the conformal transformation on the flat space-time, is a symmetry of the system.

The contraposition of Zumino's theorem says that if the theories are not invariant under the full Virasoro symmetry, it cannot be Weyl invariant in the general curved background. Because if it were the case, it would be Virasoro invariant from the above argument. Thus we see that the globally conformal differential equations without the Virasoro invariance, which we have studied in the previous section, cannot be uplifted to any Weyl invariant equations in general curved background. Of course, this argument alone does not tell whether the Virasoro invariant equations can be uplifted to the Weyl invariant equations, but we will show this is the case (up to possible quantum anomaly we do not talk about).

Let us take a look at some examples. 
Consider the Laplace equation 
\begin{align}
\partial^\mu \partial_\mu \phi = 0 \ 
\end{align}
in flat space-time in general dimensions.
On the general curved space-time, it can be uplifted to the so-called conformal Laplace equation
\begin{align}
(D^\mu D_\mu + \xi R) \phi = 0 \ ,
\end{align}
which is Weyl invariant, by assigning the Weyl weight $\tilde{\Delta} = \frac{d-2}{2}$ to $\phi$ and setting $\xi = \frac{d-2}{4(d-1)} $. By using Zumino's theorem we know that the Laplace equation is conformal invariant in any dimensions. In particular, we know that the  Laplace equation is Virasoro invariant in two-dimensions as we know a free massless boson has the Virasoro symmetry.

In contrast, let us consider the second power of Laplace equation (or dipole equation)
\begin{align}
(\partial^\mu \partial_\mu)^2 \phi = 0
\end{align}
in flat space-time. We know that it is (globally) conformal invariant in any dimensions, but it is not Virasoro invariant in two-dimensions. The Weyl invariant counterpart of this equation in general dimensions was proposed by Paneitz \cite{PA} and it is given by
\begin{align}
(D^2)^2\phi+ D_\mu \left(-\frac{4}{d-2} R^{\mu\nu} + \frac{d^2-4d+8}{2(d-1)(d-2)}g^{\mu\nu} R \right) \partial_\nu \phi \cr
+ (d-4)\left(\frac{1}{4(d-1)}D^2 R - \frac{1}{(d-2)^2} R^{\mu\nu} R_{\mu\nu}+ \frac{d^3-4d^2+16d-16}{16(d-1)^2(d-2)^2}\right)\phi = 0 \ .
\end{align}
The Paneitz equation\footnote{In four dimensions, the same equation was studied by Fradkin-Tseytlin \cite{Fradkin:1982xc}\cite{Fradkin:1981jc} and by Riegert \cite{Riegert:1984kt} independently.} does not make sense in two-dimensions, and we may conclude that the second power of the Laplace equation in two-dimensions cannot be made Weyl invariant in the Paneitz way, but is there really no other way? The argument based on Zumino's theorem above states that indeed there is no other way. Otherwise, the second power of the Laplace equation should be Virasoro invariant in flat two-dimensional space-time, but it is not.
 Similarly, the higher conformal powers of Laplacian (known as GJMS operators \cite{GG}) do not exist in two-dimensions, in agreement with our discussions.

On the other hand, our studies  in the previous section show that there may exist Weyl invariant first order differential equations in two-dimensions. These are the only equations that can be compatible with the Virasoro symmetry from  Zumino's theorem if any. Indeed, let us consider the spin $l$ conserved current in the curved background
\begin{align}
D_{\mu_1} J^{\mu_1\mu_2\cdots \mu_l} = 0 \ , \label{hsc}
\end{align}
where $J$ is the traceless symmetric tensor with conformal weight $\Delta = l$. Vanishing of the Weyl variation under the Weyl transformation of $g_{\mu\nu} \to \Omega^2 g_{\mu\nu}$ and $ J^{\mu_1\mu_2\cdots \mu_l} \to \Omega^s J^{\mu_1\mu_2\cdots \mu_l}$ requires 
\begin{align}
\Omega^s D_{\mu_1} J^{\mu_1\mu_2\cdots \mu_l} + (s + d +2(l-1))\Omega^{s-1}(D_{\mu_1} \Omega )J^{\mu_1\mu_2\cdots \mu_l} = 0 
\end{align}
by using the crucial traceless condition on $ J^{\mu_1\mu_2\cdots \mu_l}$. Thus by choosing $s= - 2l$ in $d=2$, the equation is Weyl invariant. The conformal weight is then given by the formula $\Delta = -s - l = l$. We therefore showed the existence of the Weyl invariant equations corresponding to the spin $l$ conserved current.

There are dual equations associated with the higher spin conserved current given by
\begin{align}
P_s( D_\mu J_{\mu_1 \mu_2 \cdots \mu_l}) = 0
\end{align}
where with $P_s$ we symmetrize and subtract the trace in the entire indices. 
These are dual in the sense that it is obtained from the Weyl invariant effective action
\begin{align}
S =  \int d^2x \sqrt{g} O_{\mu_1 \mu_2 \cdots \mu_l} D_{\mu} J^{\mu \mu_1 \mu_2 \cdots \mu_l} \ 
\end{align}
by varying the symmetric traceless tensor $J^{\mu_1 \mu_2 \cdots \mu_l}$. If we varied the other symmetric traceless tensor $O_{\mu \mu_1  \cdots \mu_l}$ instead, we would get the conservation equation \eqref{hsc}. These equations are known as conformal Killing tensor equations. The existence of the Weyl invariant effective action ensures that the conformal Killing tensor equations are Weyl invariant.

We have presented all the Weyl invariant differential equations that can be constructed out of single Verma module. 
Our discussions show that any other equations cannot be Weyl invariant, but it is instructive to see what goes wrong with them. Suppose we consider the (maximally) partially conserved current of the form
\begin{align}
\partial_{\mu_1} \cdots  \partial_{\mu_l} J^{\mu_1 \cdots \mu_l} = 0 \ .
\end{align}
This equation is (globally) conformal invariant with the conformal weight $\Delta = 1 $. 

In general dimensions $d$, one may find the explicit form of the  Weyl invariant uplift for the small $l$:
\begin{align}
D_\mu J^\mu &= 0 \cr
(D_{\mu_1} D_{\mu_2} + \frac{1}{d-2} R_{\mu_1 \mu_2}) J^{\mu_1 \mu_2} &= 0 \cr
(D_{\mu_1} D_{\mu_2} D_{\mu_3} + \frac{4}{d-2} R_{\mu_1 \mu_2} D_{\mu_3} + \frac{2}{d-2} (D_{\mu_1} R_{\mu_2 \mu_3})) J^{\mu_1 \mu_2 \mu_3} & = 0 \ .
\end{align}
More generally, we conjecture
\begin{align}
\left(D_{\mu_1} D_{\mu_2} \cdots D_{\mu_l} + \frac{l(l-1)(l+1)}{6(d-2)} R_{\mu_1 \mu_2} (D_{\mu_3}\cdots D_{\mu_l}) + \cdots \right)J^{\mu_1 \mu_2 \cdots \mu_l} =0 
\end{align}
with the higher curvature terms to be determined. We see that these expression does not make sense in $d=2$ and the Weyl invariant partially conserved current does not exist. The explicit form up to $l=8$ can be found in \cite{CDS}, and it was shown that there does not exist any obstructions in $d>2$ for general $l$.

The other examples we would like to mention is the conformal Laplace equations on spin $l$ symmetric tensors. In general dimension $d$, the Weyl invariant Laplace-like equations was presented in \cite{Erdmenger:1997wy}
\begin{align}
0 =& D^2 J_{\mu_1 \cdots \mu_l}  -\frac{4l}{d+2l-2} D_{\mu_l}D^{\lambda} J_{\mu_1 \cdots \mu_{l-1} \lambda}  \cr
& -\frac{d-2}{4(d-1)} R J_{\mu_1 \cdots \mu_l} -\frac{2l}{d-2} (R_{\mu_l}^{\lambda} -\frac{R}{2(d-1)} \delta_{\mu_l}^{\lambda}) J_{\mu_1 \cdots \mu_{l-1} \lambda}   
\end{align}
which, however, does not exist in $d=2$ dimensions, in agreement with our results.\footnote{To be more precise, in two-dimensions, the last term of $(R_{\mu}^{\nu} -\frac{R}{2(d-1)} \delta_{\mu}^{\nu})$ vanishes, which would be the only term that gives the Weyl variation of $D^\mu D_\nu \log\Omega$ in the other dimensions. In other words, in two-dimensions, there is no way to cancel the Weyl variation of the form $D^\mu D_\nu \log\Omega$ that comes from the first two terms in \cite{Erdmenger:1997wy}. The author would like to thank H.~Osborn for asking him about the potential issue of the limit here.} 
The non-existence of the Weyl invariant equations for $l=1$ case was also mentioned in \cite{Okui:2018oxl}.\footnote{Branson constructed second powers of conformal Laplacian on $k$-forms except in $d=2,4$ \cite{math2}. The impossibility of the  $k=1,2$ cases in $d=2$ is relevant for us.}

\section{Discussions}
In recent years, there has been an interest in deriving bounds on conformal data in conformal field theories by using a method of numerical conformal bootstrap only based on the global conformal symmetry rather than the full Virasoro symmetry. It has been surprising to see how the global conformal invariance alone can spot some of the non-trivial fixed points in two-dimensional conformal field theories. In deriving these constraints, it is important to understand the pole structures in conformal blocks, which we give physical interpretations based on globally conformal invariant differential equations. We found that almost all the globally conformal invariant differential equations in two dimensions are neither Virasoro invariant nor Weyl invariant. The only exceptions are the higher spin conservation laws, conformal Killing tensor equations and the Laplace equation of a conformal scalar. This may imply some obstructions to set up (global) conformal bootstrap analysis in more non-trivial backgrounds.

One possible future direction is a study of the similar obstructions in higher dimensions.
As far as the author is aware, the explicit construction of (all the) Weyl invariant differential equations in general curved background has not been completed yet. The naive expectation was that for each conformal invariant operators in flat space-time, there should exist a corresponding Weyl invariant operator, but as we have seen, there has been obstructions (see \cite{G}\cite{GH} for mathematical studies in higher dimensions). In two-dimensions, the obstructions can be physically understood as the lack of Virasoro invariance from Zumino's theorem. Although there is no Virasoro symmetry in higher dimensions, it would be interesting if the physical argument can provide some hints toward the fundamental mathematical understanding of conformal invariant differential operators.\footnote{Obviously, there are many related works in mathematics including \cite{ES}. The obstructions in powers of Laplacian is deeply connected with the Graham-Fefferman expansions of ambient metric and may be related to the AdS/CFT correspondence.}

Another physical signification of the non-existence of Weyl invariant differential operators is the Weyl anomaly of the conformal anomaly. Suppose we have a  scalar operator $O$ with $\Delta = 3$ in two-dimensional conformal field theories. Let $\lambda(x)$ be the corresponding source. Then the conformal anomaly may contain the term
\begin{align}
\delta Z  = \int d^2x \sigma(x) \lambda (x) \partial^2 \bar{\partial}^2 \lambda(x) \ 
\end{align}
under the infinitesimal Weyl variation $\Omega(x) \sim 1 + \sigma(x)$.
However, there is no Weyl invariant uplift in the curved background (unlike in $d=4$ dimensions), so the Wess-Zumino consistency condition becomes much more complicated, which has been studied very recently in \cite{Schwimmer:2019efk}.\footnote{A related studies (in the regime where the subtlety did not arise) can be found in \cite{Farnsworth:2017tbz}.} The similar obstruction appears in the $T\bar{J}$ deformations \cite{Guica:2017lia}\cite{Bzowski:2018pcy}\cite{Chakraborty:2018vja}, which gives
\begin{align}
\delta Z  = \int d^2x \sigma(x) \lambda (x) \partial \bar{\partial}^3 \lambda(x) \ ,
\end{align}
and again our discussions suggest the conformal anomaly does not have the Weyl invariant uplift in the curved background. This anomaly vanishes in the very special $T\bar{J}$ deformations studied in \cite{Nakayama:2018ujt} because $\bar{J}$ is a ``null" current there, but generically it is non-vanishing, which implies a non-trivial renormalization group structure in the curved background.

\section*{Acknowledgements}
This work is in part supported by JSPS KAKENHI Grant Number 17K14301. 
The author would like to thank M.~Eastwood, C.~Fefferman and R.~Graham for guiding him by providing relevant literature in mathematics.

\appendix 

\section{Weyl invariance with twisted energy-momentum tensor}
In this appendix, we discuss obstructions for the Weyl invariance due to a gravitational anomaly. 
Suppose we have a conformal field theory with $U(1)$ current algebra generated by the conserved traceless energy-momentum tensor $T$ and the $U(1)$ current $J$. Here, we take the canonical energy-momentum tensor $T$ (e.g. constructed by the Sugawara form). 

This theory admits a one-parameter deformation of the energy-momentum tensor  given by the twist
\begin{align}
T' = T + s \partial J \ .
\end{align}
The twisted energy-momentum tensor is still conserved and traceless $\bar{\partial} {T'} = 0$ so that it generates the twisted Virasoro symmetry. Indeed $L'_{n} = L_n +  s (n+1) J_n$ satisfies the Virasoro algebra (with different central charge). Note that without changing the Hilbert space, the original representation of the Virasoro algebra (generically) does not show the unitary representation of the twisted Virasoro algebra,\footnote{In particular, $J$ is no longer primary nor descendant, which is impossible in unitary conformal field theories \cite{Friedan:1985ge}\cite{Nakayama:2018dig}. This difficulty was also studied in mathematical literatures  \cite{Guido:1997gi}\cite{Morinelli:2018pof}. The author would like to thank Y.~Tanimoto for discussions and comments.} but this will not be critical in the following.

Can we construct the Weyl invariant uplift with the twisted energy-momentum tensor? It is not always the case. Let us take an example of free compact boson $X$ with a radius $r$ (i.e. $X \sim X+ 2\pi r$) with the $U(1)$ current $J = \partial X$. In order to realize the twisted energy-momentum tensor
\begin{align}
T' = -\frac{1}{2} \partial X \partial X + s \partial^2 X 
\end{align}
in the curved background, we need to add the curvature coupling 
\begin{align}
\delta S  = \int d^2x \sqrt{g} s X R \ .
\end{align}
However, this curvature coupling does not respect the identification $X \to X + 2\pi r$ unless $s$ is pure imaginary and quantized in the unit of $1/r$. 

Therefore this theory with generic $s$ does not admit a Weyl invariant uplift, or more precisely there is no diffeomorphism invariant uplift in the curved background, so it evades the argument by Zumino's theorem in the main text. 
One may trace back the difficulty to the gravitational anomaly: the twisted energy-momentum tensor has the spectrum with non-integer spin $L_{0}' - \bar{L}_0'$. Only when $s$ takes particular values, the twisted spin becomes integers and the theory can be put on the general curved background (at the sacrifice of unitarity).
 
Note that when $X$ is non-compact, the above difficulty does not arise because the twisting does not change the spin. The theory is known as linear dilaton theory and it has the Weyl invariant uplift. Another possibility is to restrict the spectrum so that the $L_{0}'-\bar{L}_0'$ is an integer: this case was covered by the Coulomb gas approach to minimal models.

\end{document}